# REAL-TIME AND SEAMLESS MONITORING OF GROUND-LEVEL PM$_{2.5}$ USING SATELLITE REMOTE SENSING

Tongwen Li [1], Chengyue Zhang [1], Huanfeng Shen [1, 4, *], Qiangqiang Yuan [2, 4], Liangpei Zhang [3, 4]

[1] School of Resource and Environmental Sciences, Wuhan University, Wuhan, China - litw@whu.edu.cn, cowherbseed@gmail.com, shenhf@whu.edu.cn
[2] School of Geodesy and Geomatics, Wuhan University, Wuhan, China - yqiang86@gmail.com
[3] The State Key Laboratory of Information Engineering in Surveying, Mapping and Remote Sensing, Wuhan University, Wuhan, China - zlp62@whu.edu.cn
[4] The Collaborative Innovation Center for Geospatial Technology, Wuhan, China

**Commission III, WG III /8**

**ABSTRACT:**

Satellite remote sensing has been reported to be a promising approach for the monitoring of atmospheric PM$_{2.5}$. However, the satellite-based monitoring of ground-level PM$_{2.5}$ is still challenging. First, the previously used polar-orbiting satellite observations, which can be usually acquired only once per day, are hard to monitor PM$_{2.5}$ in real time. Second, many data gaps exist in satellite-derived PM$_{2.5}$ due to the cloud contamination. In this paper, the hourly geostationary satellite (i.e., Himawari-8) observations were adopted for the real-time monitoring of PM$_{2.5}$ in a deep learning architecture. On this basis, the satellite-derived PM$_{2.5}$ in conjunction with ground PM$_{2.5}$ measurements are incorporated into a spatio-temporal fusion model to fill the data gaps. Using Wuhan Urban Agglomeration as an example, we have successfully derived the real-time and seamless PM$_{2.5}$ distributions. The results demonstrate that Himawari-8 satellite-based deep learning model achieves a satisfactory performance (out-of-sample cross-validation R$^2$=0.80, RMSE=17.49 $\mu g/m^3$) for the estimation of PM$_{2.5}$. The missing data in satellite-derive PM$_{2.5}$ are accurately recovered, with R$^2$ between recoveries and ground measurements of 0.75. Overall, this study has inherently provided an effective strategy for the real-time and seamless monitoring of ground-level PM$_{2.5}$.

## 1. INTRODUCTION

Previous studies have indicated that long-term exposure to PM$_{2.5}$ (airborne particles with aerodynamic diameter of less than 2.5 $\mu m$) is associated with many health concerns, such as cardiovascular and respiratory morbidity and mortality (Madrigano et al., 2013). However, the assessment of PM$_{2.5}$ exposure is limited due to the sparse and uneven distribution of ground monitoring stations.

Satellite remote sensing has the potential to expand PM$_{2.5}$ estimation beyond those only provided by ground stations (Li et al., 2016). The most widely used satellite parameter is aerosol optical depth (AOD) (Hoff and Christopher, 2009). Many satellite instruments own the capacity to provide AOD products, and have been applied to the monitoring of PM$_{2.5}$, for instance, the Moderate Resolution Imaging Spectroradiometer (Li et al., 2017b) and Multiangle Imaging SpectroRadiometer (You et al., 2015) on board Earth Observing System (EOS) satellites (i.e., Terra and Aqua) etc.

However, previous satellite-based PM$_{2.5}$ estimation usually rely on polar-orbiting satellites (e.g., Terra, Aqua). In general, this type of satellite provides only one observation per day for a given location. This means the polar-orbiting satellite will not be able to monitor PM$_{2.5}$ once again until the next day. Hence, the PM$_{2.5}$ pollution (especially sudden pollution event) may not be monitored in real time by polar-orbiting satellite.

Furthermore, previous studies have suggested that there exist diurnal variation of PM$_{2.5}$ (Guo et al., 2016). The diurnal variation of PM$_{2.5}$ cannot be effectively characterized by the polar-orbiting satellites. With a high temporal resolution (e.g., 1 hour), the geostationary satellite has been attempted to be used for the estimation of ground PM$_{2.5}$/PM$_{10}$. The results indicated that the hourly geostationary observations show great potential in the real-time monitoring of PM$_{2.5}$/PM$_{10}$ (Emili et al., 2010; Paciorek et al., 2008). While these studies still mainly focused on PM$_{2.5}$/PM$_{10}$ estimation at a daily scale, and the hourly estimation accuracy has great room for improvement.

On the other hand, due to the cloud contamination, there are many gaps in satellite remote sensing data (Li et al., 2014; Shen et al., 2014). The satellite-based PM$_{2.5}$ estimates are usually seamed in space. To address this issue, two main strategies have been carried out. Firstly, the satellite AOD products were spatially interpolated to improve its coverage (Lv et al., 2016; Ma et al., 2014). Secondly, the spatial smooth techniques were adopted to fill the missing data of satellite-derived PM$_{2.5}$ (Just et al., 2015; Kloog et al., 2011). For these two strategies, they mainly considered spatial correlation information of aerosol/PM$_{2.5}$ for the reconstruction of missing data. There may be great uncertainty, especially for large gaps, because of the lack of preference data. It should be noted that the valid observations at a nearby time may exist, and they are a good supplementary for the reconstruction of PM$_{2.5}$. Whether is it possible to fuse the spatial and temporal correlation information

---

* Corresponding author, Huanfeng Shen, Email: shenhf@whu.edu.cn

of satellite and ground observations for the reconstruction of missing PM$_{2.5}$?

Here arrives at the objective of this study. Firstly, taking advantage of geostationary satellite observations with high temporal resolutions, we would like to improve the time efficiency of satellite-based monitoring of PM$_{2.5}$ from the daily scale to the hourly scale. In addition, a spatio-temporal fusion technique is developed to recover the missing data of satellite-derived of PM$_{2.5}$ using time series observations of satellite and ground stations. Therefore, the purpose of this paper is to derive hourly and seamless PM$_{2.5}$ distributions by fusing the satellite remote sensing and ground station measurements.

## 2. STUDY REGION AND DATA

### 2.1 Study region and period

The study region is Wuhan Urban Agglomeration (WUA), which is presented in Figure 1. The study period is a total year of 2016. WUA is located in central China (as shown in Figure 1). To make full use of PM$_{2.5}$ station measurements, the monitors in the range with latitude of 28.4°~32.3°N and longitude of 112.0°~116.7°E are all included in our analysis. WUA is an urban group with the center of Wuhan, covering the vicinal 8 cities (Huangshi, Ezhou, Huanggang, Xiaogan, Xianning, Xiantao, Qianjiang, and Tianmen).

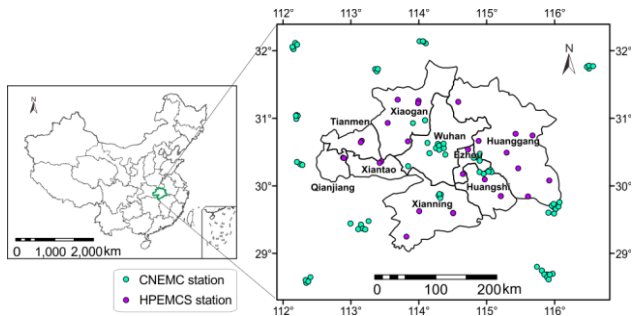

Figure 1. Study region and spatial distribution of PM$_{2.5}$ stations.

### 2.2 Ground-level PM$_{2.5}$ measurements

We collected hourly PM$_{2.5}$ data from the China National Environmental Monitoring Center (CNEMC) website (http://www.cnemc.cn) and the Hubei Provincial Environmental Monitoring Center Station (HPEMCS) website (http://www.hbemc.com.cn/). In this study, 77 CNEMC stations and 27 HPEMCS stations (104 stations in total) are included. The distribution of PM$_{2.5}$ stations is shown in Figure 1.

### 2.3 Himawari AOD

Himawari-8 is one of the third generation of geostationary weather satellites, launched on 7th October 2014 carrying the new AHI instrument. The Himawari-8 has an observation range of 80°E ~ 160°W and 60°N ~ 60°S, with the center of 140.7°E over equator. The aerosol optical depth product is derived from Himawari-8 visible and near-infrared data. This product provides information on AOD at 500 nm for areas over oceans and land during the daytime. The algorithm references a look-up table calculated on the basis of an assumed spheroid-particle aerosol model (Fukuda et al., 2013).

The Himawari-8 Level 3 hourly AOD data corresponding to the ground-level PM$_{2.5}$ measurements were downloaded from Japan Aerospace Exploration Agency (JAXA) P-Tree System (http://www.eorc.jaxa.jp/ptree/). This AOD products have a spatial resolution of 5 km, and they are available every 1 hour. In this study, only aerosol retrievals with the highest confidence level ("very good") were adopted for the estimation of PM$_{2.5}$.

### 2.4 Meteorological parameters and land cover data

The Goddard Earth Observing System Data Assimilation System GEOS-5 Forward Processing (GEOS 5-FP) meteorological data were used in this study. The reanalysis meteorological data have a spatial resolution of 0.25° latitude × 0.3125° longitude. Hourly specific humidity (SH, kg/kg), air temperature at a 2 m height (TMP, K), wind speed at 10 m above ground (WS, m/s), and, surface pressure (PS, kPa) from this datasets were extracted. Each variable was regridded to 0.05° to be consist with satellite observations. More details about the GEOS 5-FP data can be found at the website (https://gmao.gsfc.nasa.gov/forecasts/).

MODIS normalized difference vegetation index (NDVI) products (MOD13) were downloaded from the NASA website (https://ladsweb.modaps.eosdis.nasa.gov/). This product is available at a resolution of 1 km every 16 days, and was incorporated to reflect the land-use type.

### 2.5 Data pre-processing and matching

Firstly, we created a 0.05-degree grid for the data integration, model establishment, and spatial mapping. For each 0.05-degree grid, ground-level PM$_{2.5}$ measurements from multiple stations are averaged. Meanwhile, we resampled the meteorological data to match with satellite observations. All the data were re-projected to the same coordinate system. Finally, we extracted satellite observations, meteorological parameters on the locations where PM$_{2.5}$ measurement are available.

## 3. METHODOLOGY

The main procedure of our method includes two parts, which is illustrated in Figure 2. Firstly, a deep learning architecture is developed to estimate ground-level PM$_{2.5}$ using Himawari AOD and auxiliary predictors. On this basis, the satellite-derived PM$_{2.5}$ in conjunction with ground-level PM$_{2.5}$ measurements are incorporated into a spatio-temporal fusion model for the reconstruction of PM$_{2.5}$. The details of each part can be seen in Section 3.1 and 3.2.

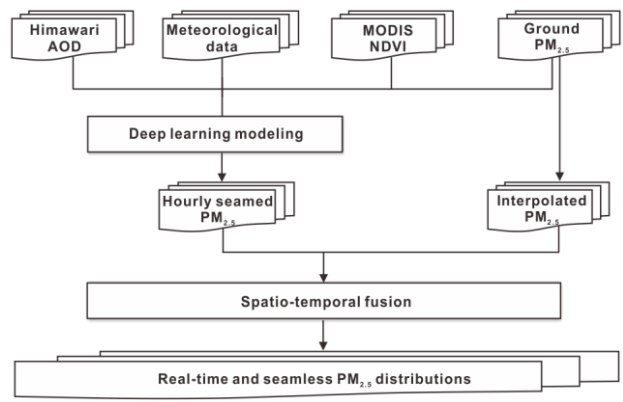

Figure 2. Flowchart describing the procedure for deriving hourly and seamless PM$_{2.5}$.

## 3.1 Deep learning for the satellite-based PM$_{2.5}$ estimation

The deep belief network (DBN) model, which is one of the most typical deep learning models (Hinton et al., 2006), was introduced to represent the relationship between PM$_{2.5}$, AOD, and auxiliary factors. Additionally, the geographical correlation of PM$_{2.5}$ were incorporated into the DBN model (Geoi-DBN) (Li et al., 2017a). Because the nearby PM$_{2.5}$ from neighbouring stations and the PM$_{2.5}$ observations from prior days for the same station are informative for estimating PM$_{2.5}$. The general structure of Geoi-DBN model used to estimate PM$_{2.5}$ is:

$$PM_{2.5} = f(AOD, SH, WS, TMP, PS, NDVI, S\text{-}PM_{2.5}, T\text{-}PM_{2.5}, DIS) \quad (1)$$

where $S\text{-}PM_{2.5}, T\text{-}PM_{2.5}, DIS$ denote as the geographical correlation of PM$_{2.5}$, their calculation can be found at Li et al. (2017a). A Geoi-DBN model comprising two restricted Boltzmann machine (RBM) layers for estimating ground-level PM$_{2.5}$ is presented in Figure 3. The input variables are satellite AOD, meteorological parameters, NDVI, and geophysical correlation of PM$_{2.5}$; the output is ground PM$_{2.5}$. The Geoi-DBN model is firstly trained using the collected AOD-PM$_{2.5}$ matchups, and subsequently utilized to predict spatial values where there are no monitoring stations.

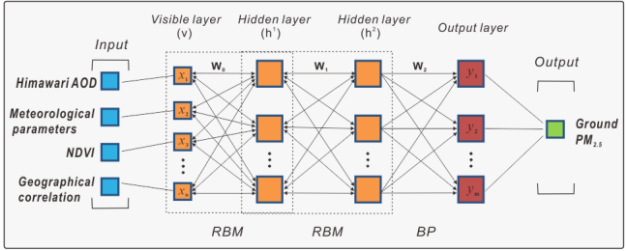

Figure 3. The structure of Geoi-DBN for PM$_{2.5}$ estimation.

## 3.2 Spatio-temporal fusion for the reconstruction of PM$_{2.5}$

The optical satellite is often impacted by clouds, and thus the spatial discontinuity exists in satellite-derived PM$_{2.5}$. To address this issue, using time series satellite-retrieved PM$_{2.5}$ and station-measured PM$_{2.5}$, we propose a satellite-station spatio-temporal fusion method to fill the data gaps. The initial PM$_{2.5}$ data are obtained by the interpolation of station PM$_{2.5}$ using the inverse distance weighting (IDW) method. The basic supposition is that the variation of the interpolated PM$_{2.5}$ remains similar trend with that of the satellite-derived PM$_{2.5}$ between the same periods for a given location.

For convenience, we refer to the interpolated PM$_{2.5}$ as coarse-resolution data, and the satellite-derived PM$_{2.5}$ (in Section 3.1) as fine-resolution data. Then, we denote the given seamed PM$_{2.5}$ as target data. To reconstruct the target data, the auxiliary data used are $N$ pairs of coarse- and fine-resolution data acquired prior the target time $T_p$ and the coarse-resolution data at the target time. For a given missing pixel:

$$PM(x, y, T_p) = \sum_k W(T_k) \left( \sum_{i=1}^{n} W(x_i, y_i, T_k) \times [a(x, y, \Delta T_k) * PM(x_i, y_i, T_k) + b(x, y, \Delta T_k)] \right) \quad (2)$$

where $PM(x, y, T_p)$ is the prediction of the missing pixel $(x, y)$ at prediction time $T_p$; $n$ is the number of similar pixels (with similar PM$_{2.5}$ values and spatial patterns) for the missing pixel; $(x_i, y_i)$ is the location of the $ith$ similar pixel, and $W(x_i, y_i, T_k)$ is the weight of $ith$ similar pixel on the auxiliary fine-resolution data at time $T_k$. The selection of similar pixels and the calculation of their weights refers to Cheng et al. (2017). $a(x, y, \Delta T_k)$ and $b(x, y, \Delta T_k)$ are regression coefficients fitted using the similar pixels on the coarse-resolution data at time $T_p$ and $T_k$, and they are transferred to the fine-resolution data for the prediction of missing data.

On the other hand, we can obtain $N'$ ($N' \leq N$) (due to the missing data on auxiliary data) predictions for the missing pixels using $N$ pairs of auxiliary data. To reduce the random error, the weighted average of them is considered as the final prediction. The weights ($W(T_k)$) are set according to PM$_{2.5}$ variation between prediction time ($T_p$) and auxiliary time ($T_k$):

$$W(T_k) = \frac{1}{\sum_{k=1}^{N'} \frac{1}{RMSE_{\Delta k}}} \left( \frac{1}{RMSE_{\Delta k}} \right) \quad (3)$$

where $RMSE_{\Delta k}$ means root-mean-square error between coarse data at prediction time ($T_p$) and auxiliary time ($T_k$). Here, the "spatio-temporal fusion" means the combination of spatial information (similar pixels) and temporal information (data at auxiliary time) for the reconstruction of PM$_{2.5}$.

Through the above reconstruction process, some tiny regions are still missing due to the lack of auxiliary data. They are interpolated using the IDW method to achieve full-coverage PM$_{2.5}$ data.

## 3.3 Model evaluation

Firstly, to evaluate the accuracy of PM$_{2.5}$ retrieval, a 10-fold cross-validation technique (Rodriguez et al., 2010) was adopted to test the potential of model overfitting for Geoi-DBN. All samples in the model dataset are randomly and equally divided into ten subsets. One subset is used as validation samples and the rest subsets are used to fit the model for each round of validation. We adopted the coefficient of determination ($R^2$), the root-mean-square error (RMSE, $\mu g/m^3$), the mean prediction error (MPE, $\mu g/m^3$), and the relative prediction error (RPE, defined as RMSE divided by the mean ground-level PM$_{2.5}$) to evaluate the model performance.

Secondly, in order to verify the accuracy of PM$_{2.5}$ reconstruction, we compared the reconstruction PM$_{2.5}$ in a total year of 2016 with the corresponding station measurements. Statistical indices of the $R^2$ and RMSE are used to give a quantitative assessment.

## 4. RESULTS

### 4.1 Model evaluation

**4.1.1 Evaluation of PM$_{2.5}$ retrieval performance:** Figure 4 shows the scatter plots for the modelling fitting and cross-validation results of Geoi-DBN model. For the model fitting, the $R^2$ value is 0.80, and the RMSE is 17.37 $\mu g/m^3$. The results indicate that the Geoi-DBN model can effectively describe the AOD-PM$_{2.5}$ relationship. From model fitting to cross-validation, the $R^2$ value is equal and the RMSE only increase 0.12 $\mu g/m^3$. These findings show that the Geoi-DBN model is not over-fitted. On the other hand, the cross-validation slope of observed PM$_{2.5}$ versus estimated PM$_{2.5}$ is 0.79, reporting some evidences for bias. This means that the Geoi-DBN model tends to underestimate PM$_{2.5}$ concentrations when the ground measurements are greater than ~55 $\mu g/m^3$. The possible reason could be that we used point-based monitoring data and a spatially averaged modelling framework. The sampling distribution of monitors in a grid may not give a great estimation of the spatially averaged concentration for that grid. Generally, the Geoi-DBN model has achieved a satisfactory performance for the Himawari-based AOD estimation.

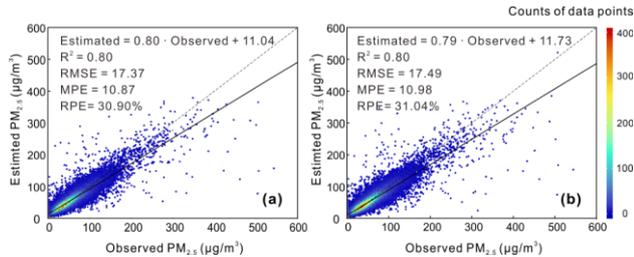

Figure 4. Scatter plots of Geoi-DBN for PM$_{2.5}$ estimation: (a) model fitting, (b) cross-validation.

**4.1.2 Evaluation of PM$_{2.5}$ reconstruction accuracy:** To evaluate the performance of PM$_{2.5}$ reconstruction, we compare the reconstruction results with ground station measurements. As shown in table1, the R$^2$ value between observed PM$_{2.5}$ and reconstruction PM$_{2.5}$ is 0.75, and the RMSE is 19.44 $\mu g/m^3$. The results show that the reconstruction PM$_{2.5}$ are highly consistent with the station measurements. For the satellite retrievals, they report R$^2$ and RMSE values of 0.81 and 16.96 $\mu g/m^3$ versus station measurements. These findings indicate that the reconstruction results almost obtain a same level of performance to the satellite retrievals, when comparing with the ground station observations. Therefore, we can say that the proposed approach is effective for reconstructing the seamless PM$_{2.5}$ distributions.

Table 1 performance between retrievals and reconstruction.

|  | R$^2$ | RMSE |
|---|---|---|
| Reconstruction results | 0.75 | 19.44 |
| Retrieval results | 0.81 | 16.96 |

**4.2 Hourly mapping of PM$_{2.5}$ distribution**

Figure 5 presents hourly satellite derived PM$_{2.5}$ on 28 February 2016. It can be clearly found that the satellite-derived PM$_{2.5}$ are all missing during 00:00 ~ 08:00 and 18:00 ~ 23:00. The reason for this is that the Himawari satellite, which is an optical satellite, has no capacity to detect the atmospheric parameters during night. During the daytime, the hourly PM$_{2.5}$ distributions are mapped. Compared to the polar-orbiting satellites, the Himawari satellite shows some advantages to investigate the diurnal variation of PM$_{2.5}$. For instance, the levels of PM$_{2.5}$ concentrations in Wuhan are becoming higher in the afternoon, especially at 15:00. However, it is worth noticing that there are still some limitations for monitoring the PM$_{2.5}$ patterns, due to the data missingness. For example, it is hard to capture the level of PM$_{2.5}$ in Huanggang at 17:00.

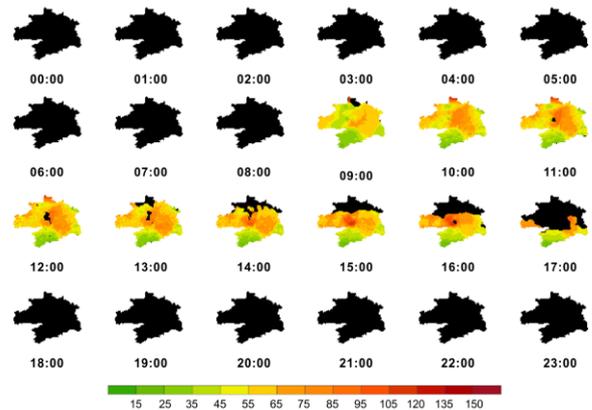

Figure 5. Hourly satellite-derived PM$_{2.5}$ on February 28, 2016. The black regions indicate missing data.

The data gaps are filled using the proposed approach, as presented in Figure 6. The missing parts of satellite-derived PM$_{2.5}$ are effectively recovered, so that the PM$_{2.5}$ patterns can be comprehensively investigated. Compared with satellite-derived PM$_{2.5}$ at 17:00 (Figure 5), we can clearly see Huanggang has a very high level of PM$_{2.5}$. Furthermore, the PM$_{2.5}$ distributions in night are also reconstructed, which cannot be directly monitored by Himawari satellite.

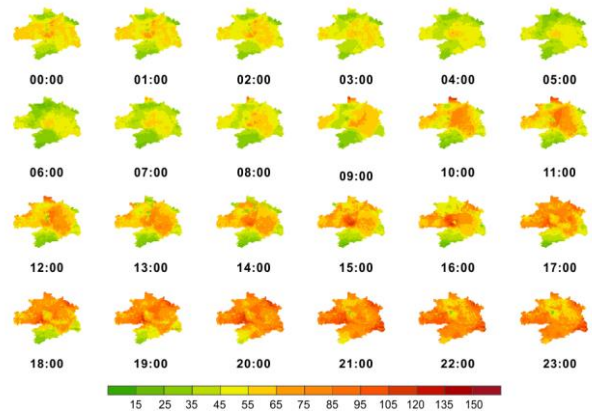

Figure 6. Hourly and seamless PM$_{2.5}$ on February 28, 2016.

## 5. CONCLUSIONS

To sum up, the hourly Himawari-8 observations are adopted to greatly improve the time efficiency of PM$_{2.5}$ monitoring. Furthermore, a spatio-temporal fusion model is applied to the fill the data gaps using satellite-derived PM$_{2.5}$ in conjunction with ground PM$_{2.5}$. The results show that Himawari-8 satellite-based deep learning model achieves a satisfactory performance (cross-validation R$^2$=0.80, RMSE=17.49 $\mu g/m^3$). The missing data in satellite-derive PM$_{2.5}$ are accurately recovered, with R$^2$ between recovery results and ground measurements of 0.75. This study has provided an effective strategy for the real-time and seamless monitoring of ground PM$_{2.5}$.

**ACKNOWLEDGEMENTS**

This work was funded by the National Key R&D Program of China (2016YFC0200900) and the National Natural Science Foundation of China (41422108).